\documentclass[review]{elsarticle}

\usepackage{lineno,hyperref}
\modulolinenumbers[5]

\journal{Optics Communications}

\begin{document}

\begin{frontmatter}

\title{Compact infrared continuous-wave double-pass\\ single-frequency doubly-resonant OPO}


\author[lac,blue]{Anne Boucon}

\author[blue]{Bertrand~Hardy-Baranski}

\author[lac]{Fabien~Bretenaker\corref{mycorrespondingauthor}}
\cortext[mycorrespondingauthor]{Corresponding author}
\ead{Fabien.Bretenaker@u-psud.fr}

\address[lac]{Laboratoire Aim\'e Cotton, CNRS-Universit\'e Paris Sud 11-ENS Cachan, Campus d'Orsay, 91405 Orsay, France}
\address[blue]{Blue Industry \& Science, 208 bis rue la Fayette, 75010 Paris, France}

\begin{abstract}
We demonstrate a compact continuous-wave single-frequency doubly-resonant optical parametric oscillator (DRO) in a double-pass pump configuration with a control of the relative phase between the reflected waves. The nested DRO cavity allows single longitudinal mode operation together with low threshold and high efficiency. Thermal effects are managed by chopping the pump beam, allowing continuous tuning of the emitted wavelength. The infrared idler wave (3200-3800 nm) can be used for gas detection and the threshold pump power is compatible with diode pumping.\end{abstract}


\end{frontmatter}


\section{Introduction}

Continuous-wave optical parametric oscillators (cw-OPOs) have unique characteristics as infrared single-frequency tunable sources \cite{Arslanov2013, Breunig2011, Hodgkinson2013} that make them extremely attractive for many applications. In particular, their narrow linewidth and high power characteristics perfectly suit sensitive and high resolution spectroscopy applications. Depending on the number of waves resonant inside the cavity (signal, idler, and/or pump), they are called singly-, doubly-, or triply-resonant oscillators.

Only few years after the first demonstration of the laser, the first realizations of cw optical parametric oscillators were obtained with a doubly-resonant cavity and with multimode pump sources \cite{Smith1968,Byer1968}. They took advantage of the fact that the threshold of doubly-resonant OPOs (DROs) is much lower than the one of singly-resonant OPOs (SROs). However, since such a DRO must obey both energy conservation and cavity resonance conditions for the two resonant waves \cite{Giordmaine1966}, one usually observes a strong sensitivity of DROs to mechanical instabilities. Moreover these oscillators are difficult to tune since the two resonant waves share the same cavity. This explains why more efforts have been devoted to SROs than to DROs when it comes to controlling the wavelength of oscillators over large wavelength tuning domains, for example for gas analysis \cite{Courtois2013}.

Nevertheless, recently, technological progress in nonlinear media and innovative cavity designs have led to renewed interest in low threshold DROs \cite{Colville1994,AlTahtamouni1998,Henderson2000,Ikegami2000}. On the one hand, the use of monolithic cavities has greatly improved the mechanical stability \cite{AlTahtamouni1998,Henderson2000,Ikegami2000}, but leaving the problem of broad frequency tunability unresolved. On the other hand,  Colville et al. \cite{Colville1994} have shown that a cw-DRO with separated cavities allows to control the frequencies of idler-signal mode pairs. Even more recently, new approaches have allowed to build monolithic and stable cavities while preserving the independent tunability of the signal and idler cavity lengths \cite{Drag2002}.  Still more recently, such a design has led to the development of the so-called nested-cavity OPO which takes advantage of double pump pass in the nonlinear crystal to further reduce the threshold.  In this design, the frequency control is improved by managing the relative phase of the three waves after reflection on the end cavity mirror \cite{Hardy2011, Barria2013}. This has been made possible by the use of a metallic mirror, instead of a dielectric mirror, as cavity rear mirror, in order to control the achromaticity of the relative phase of the three waves upon reflection on that mirror. This approach has led to excellent results in the pulsed regime. However, many applications would benefit from a cw or quasi-cw emission in order to improve the average emitted power and thus the signal-to-noise ratio. However, one must notice that the use of a metallic cavity mirror, which introduces several percents of losses for both the signal and idler, will  lead to a significant increase of the threshold with respect to a low-loss dielectric mirror. If we want to keep the OPO single-frequency and tunable, we have to pump it with a single-frequency laser. Of course, the use of a single-frequency laser diode would allow us to keep the system compact and rugged. But the maximum power of commercially available single-frequency diodes at 1064~nm is of the order of 1.5~W \cite{QPC}. The aim of the present paper is thus to investigate the cw operation of a nested-cavity DRO which must have the following characteristics: i) the rear cavity mirror is chosen to be metallic in order to make of the relative phase of the reflected pump, signal, and idler waves independent of wavelength. Together with the wedge of the output face of the nonlinear crystal, this achromatic phase upon reflection must ensure the tunability of the OPO; ii) the OPO dimensions must be kept as small as possible; iii) the threshold must be compatible with single-frequency diode pumping; iv) one must make sure that the OPO frequencies can be tuned similarly to what was obtained in pulsed regime \cite{Hardy2012}. In particular, the role of thermal effects and their impact on the frequency control of the OPO are experimentally explored.

\section{Experimental setup}

\begin{figure}[h!]
\centering
\includegraphics[width=\columnwidth]{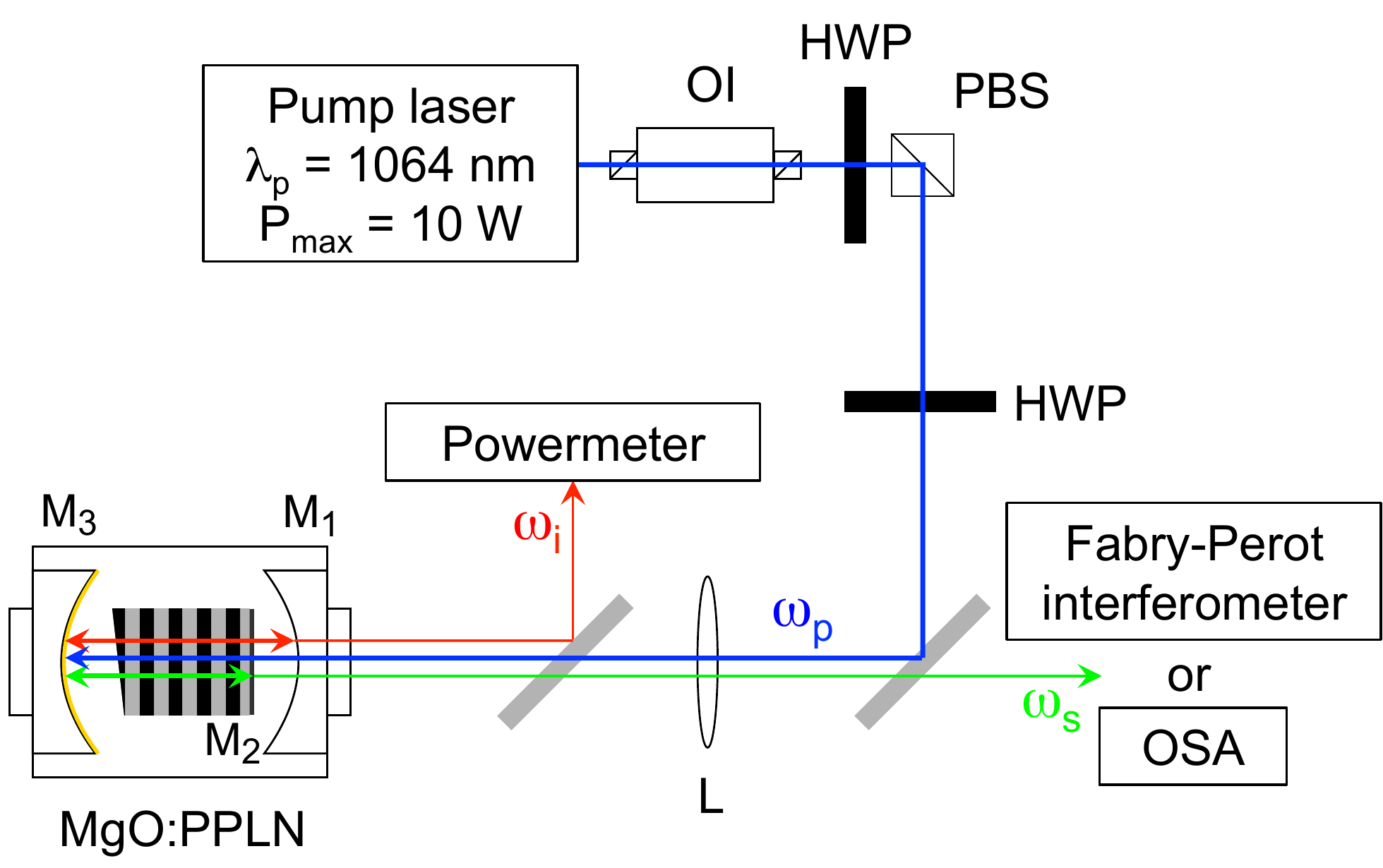}
\caption{Experimental setup. OI: optical isolator, HWP: half-wave plate, PBS: polarization beam splitter, L: focusing lens. Notice that mirror M$_3$ is metallic and reflects all the waves, including the pump. The output face of the crystal is wedged to control the relative phase of the three waves after reflection.}
\label{Fig01}
\end{figure}

The experimental setup is schematized in Fig. \ref{Fig01}. The OPO is based on a 32 mm-long magnesium oxide doped periodically poled lithium niobate (MgO-PPLN) crystal, whose temperature is stabilized at $110\,^{\circ}{\rm C}$. This crystal contains five different poling periods ranging from 29.75 to 30.75 $\mu \textrm{m}$ with a 0.25 $\mu \textrm{m}$ step. With a pump wavelength equal to 1.064~$\mu$m, this enables to cover the 3.2-3.8 $\mu \textrm{m}$ spectral range for the idler wavelength, by choosing the poling period and the temperature. 

The nested cavity is based on the use of three mirrors M$_1$, M$_2$, and M$_3$ for the signal and idler cavities \cite{Hardy2011}. The idler cavity is composed of mirrors M$_\textrm{1}$ and M$_\textrm{3}$. M$_\textrm{1}$ has a reflectivity of 97\,\% for the idler wavelength and a radius of curvature of 100 mm. Mirror M$_\textrm{3}$ is a gold mirror, which reflects 98\,\% at all the involved wavelengths, including the pump. Its radius of curvature is equal to 80~mm. The signal cavity is formed by mirror M$_3$ and mirror M$_2$ which is directly coated on the input face of the nonlinear crystal (99.8\,\% reflection between 1.48 and 1.6 $\mu \textrm{m}$). Mirrors M$_1$ and M$_3$ are located at a distance of 1 to 2 mm from the MgO-PPLN crystal. The overall length of the OPO cavity, including the thickness of the mirror mounts, is about 10 cm.

The double-pass pump configuration allows to create parametric gain in the two directions of propagation, provided the relative phase between the three waves after reflection by achromatic mirror M$_3$ is controlled \cite{Bjorkholm1970}.  To this aim, the end-face of the crystal, which is anti-reflection coated for all wavelengths, is cut with a $0.4\,^{\circ}$ angle. This wedge, together with the fact that mirror M$_3$ is achromatic, is the particularity of the present OPO design with respect to other nested cavity DROs. This idea has been initially introduced for pulsed OPOs \cite{Hardy2011}. The wedge angle permits to control the round-trip gain curve because it permits to tune the relative phase of the three waves after reflection by simply translating the crystal perpendicularly to the propagation direction. Furthermore, the fact that M$_3$ is acromatic (metallic) leads to the fact that the relative phase between the three waves is not altered upon reflection. The present work is the first investigation of this configuration in cw regime.

Mirrors M$_1$ and M$_3$ are fixed on piezoelectric transducers (PZTs). The PZT carrying M$_1$ allows to change  the idler cavity length only, while the PZT carrying M$_3$ allows to change the length of both cavities. This control permits to select only one pair of idler and signal frequencies.

The pump laser at 1.064 $\mu \textrm{m}$ is a 10~W single-frequency solid-state MOPA (Master Oscillator Power Amplifier).
After collimation and passage through an optical isolator, its power is adjusted using the combination of a half-wave plate and a polarization beam splitter. The pump beam is then focused to a 100 $\mu \textrm{m}$ waist diameter inside the OPO crystal. This pump beam size is selected to optimize the parametric conversion efficiency.

\section{CW operation characteristics}

\begin{figure}[h!]
\centering
\includegraphics[width=\columnwidth]{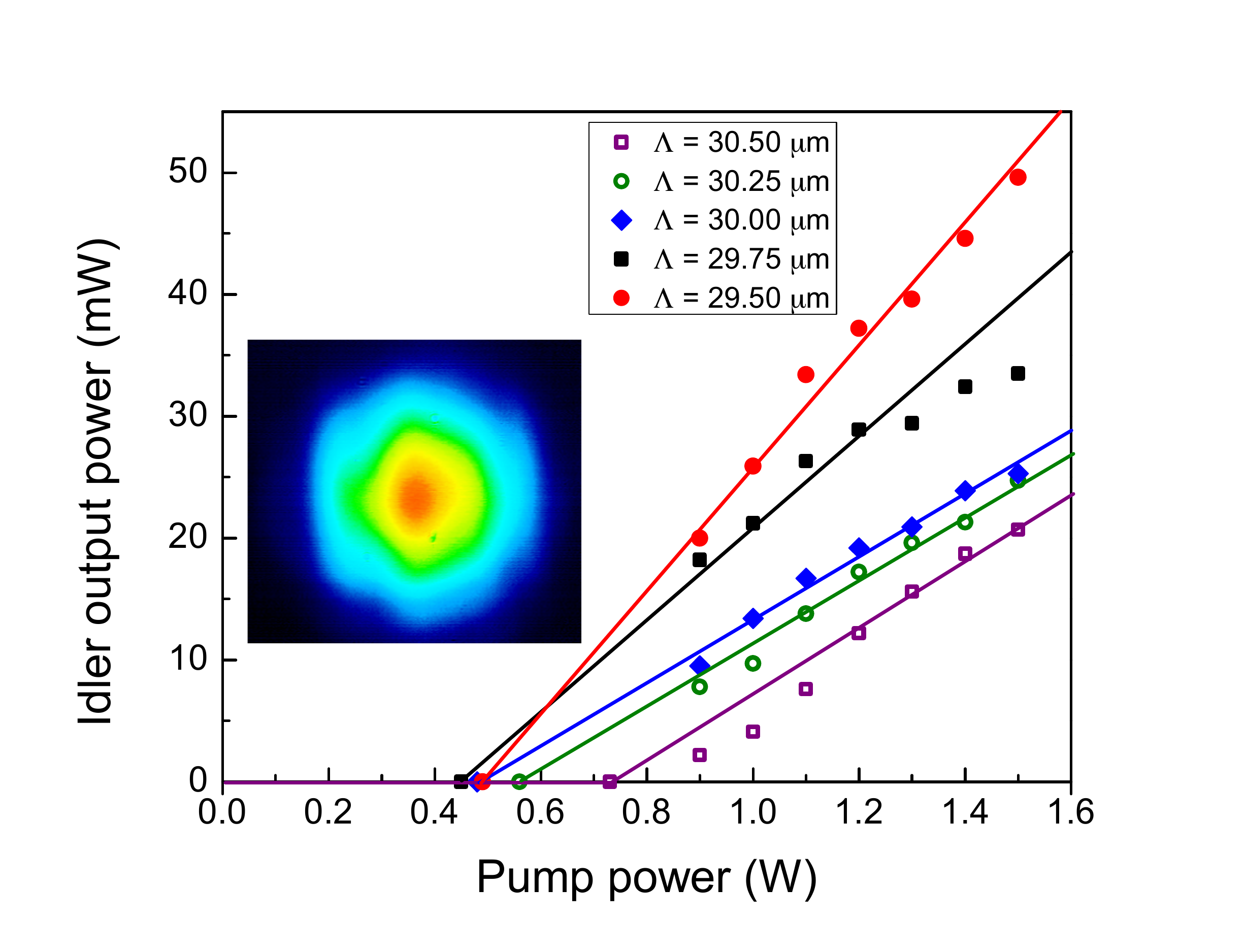}
\caption{Idler output power versus pump power for different poling periods. The lines are just guides for the eye. Measurements are performed at the output of M$_1$ and after filtering out the pump and signal beams.
Inset: idler mode profile recorded for an idler power of 15~mW. The spurious vertical fringes are due to an undesired \'etalon effect in the filter.}
\label{Fig02}
\end{figure}

Fig.\,\ref{Fig02} displays the measured output idler power versus the input pump power for five different poling periods. It is worth noticing that the threshold powers are well below 1~W, thanks to the double pump pass and doubly resonant configuration. Such threshold pump powers are compatible with the powers available from commercial single-frequency pump diodes at 1064~nm \cite{QPC}. Of course, one could argue that the threshold of a DRO could be even smaller provided one uses low loss dielectric mirror. However, here, we deliberately use a metal mirror to control the phase upon reflection. The cost of this choice is the introduction of extra losses and a not so small threshold equal to 500 mW. However, since this threshold is well below 1~W, we believe that the present architecture is a good trade-off between potential diode pumping and tunability.

Up to 50 mW idler power is achieved for a 1.5~W pump power. This efficiency is not very high, especially compared to bulkier SROs \cite{Bosenberg1996}. But it is not far from being sufficient for trace gas detections and the achieved thresholds are below 1~W, contrary to Ref. \cite{Bosenberg1996}. Moreover, although the use of a lossy metallic mirror is detrimental to the efficiency of the OPO, this output power could probably be improved by optimizing the transmission of mirror M$_1$ and the length of the nonlinear crystal. The inset in Fig.\,\ref{Fig02} shows the spatial profile of the idler beam. The fringes are due to spurious reflections inside the filter protecting the camera. At two times above threshold and a crystal temperature equal to 110 $^{\circ}$C, we measured the signal wavelengths corresponding to the poling periods $\Lambda$ equal to 29.75, 30.00, 30.25, 30. 50, and 30.75 $\mu$m. They are equal to $\lambda_{\mathrm{s}}=1513$, 1532, 1554, 1579, and 1610 nm, respectively. The corresponding idler wavelengths are $\lambda_{\mathrm{c}}=3585$, 3482, 3375, 3263, and 3189 nm, respectively.

\begin{figure}[h!]
\centering
\includegraphics[width=\columnwidth]{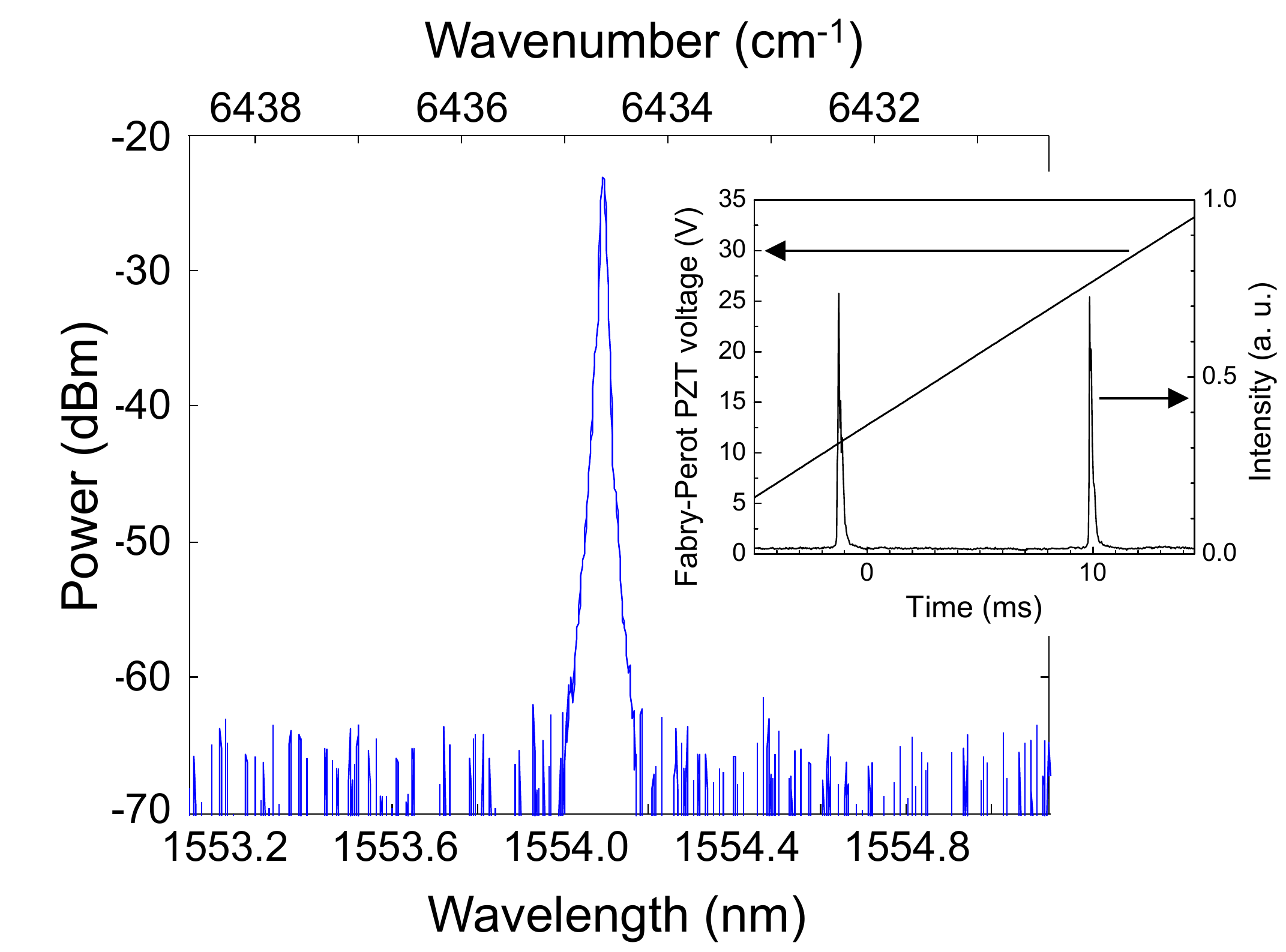}
\caption{Optical spectrum of the signal beam obtained with an optical spectrum analyzer (resolution bandwidth: 0.01~nm). Inset: Fabry-Perot spectrum (free spectral range: 1 GHz).}
\label{Fig03}
\end{figure}

We analyze the spectral purity of the OPO using an optical spectrum analyzer and a Fabry-Perot interferometer operating around 1.5 $\mu \textrm{m}$. The resulting free-running  spectra are reproduced in Fig.\,\ref{Fig03}. They show that the signal beam exhibits only one frequency. Since the pump laser is also single-frequency, this proves the single-frequency operation of the idler beam.

\begin{figure}[h!]
\centering
\includegraphics[width=\columnwidth]{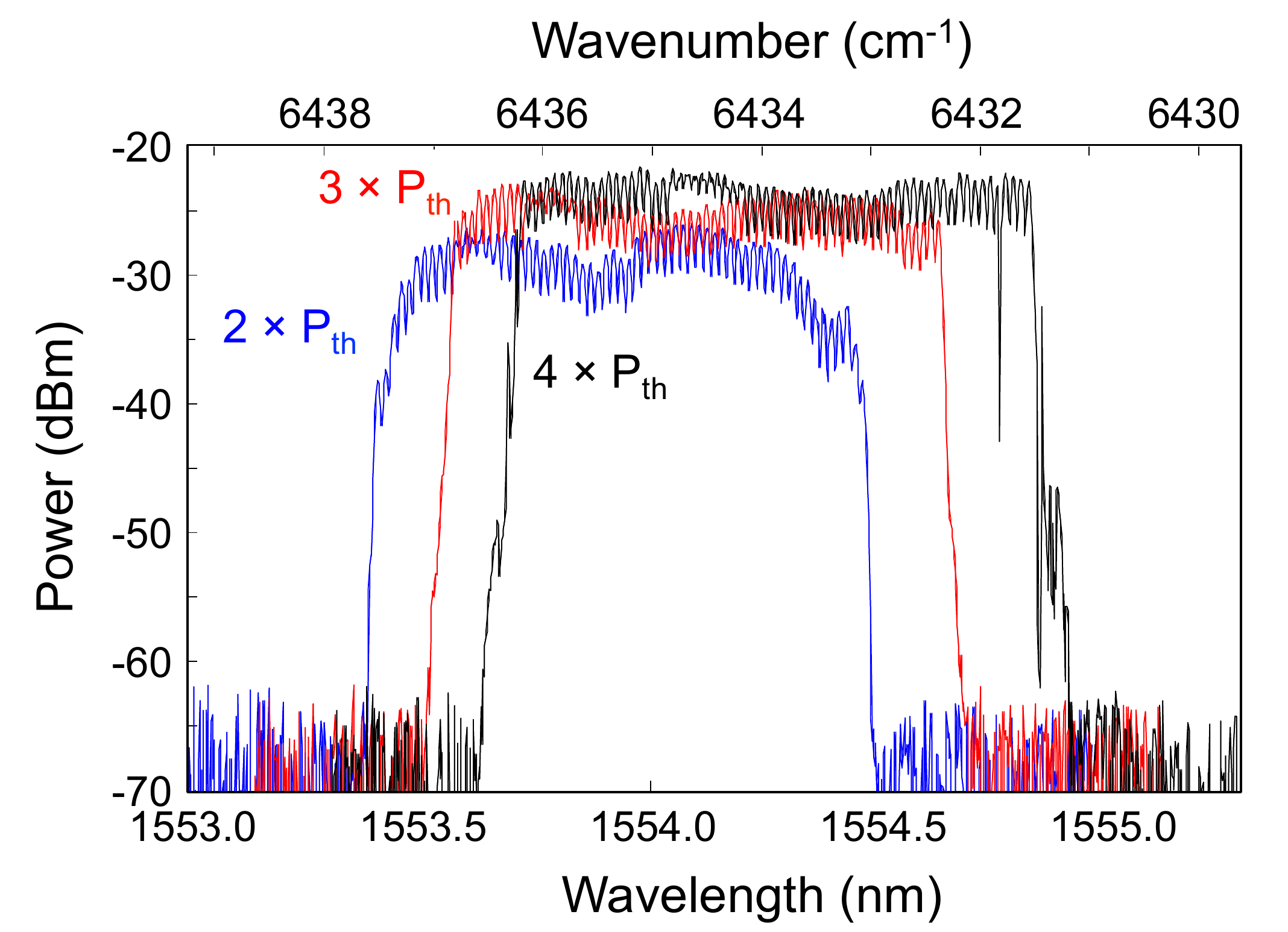}
\caption{Parametric gain bandwidth recorded for three different pumping rates. These spectra were recorded with the optical spectral analyzer in hold mode while scanning the PZTs carrying mirrors M$_1$ 	and M$_3$.}
\label{Fig04}
\end{figure}

By scanning the two cavity lengths (by sending ramp voltages on the two PZTs simultaneously), one can measure the available parametric gain bandwidth for one poling period at a given temperature ($110\,^{\circ}{\rm C}$ in the present experiment). Fig.\,\ref{Fig04} reproduces three such spectra recorded for three different pump powers (two, three, and four times above threshold) for a poling period $\Lambda = 30.25\,\mu \mathrm{m}$. One notices that the DRO can be tuned over more than 1~nm, while maintaining single-frequency operation. Besides, when the pump power is increased, the oscillation bandwidth increases also and is shifted to longer wavelengths. This shift evidences the role of thermal effects inside the crystal. Indeed, phase-matching calculations confirm that the signal wavelength increases with the crystal temperature. The red shift is compatible with the temperature rise due to the presence of the three waves inside the crystal, which modifies the refractive index and the poling period of the crystal.

\begin{figure}[h!]
\centering
\includegraphics[width=\columnwidth]{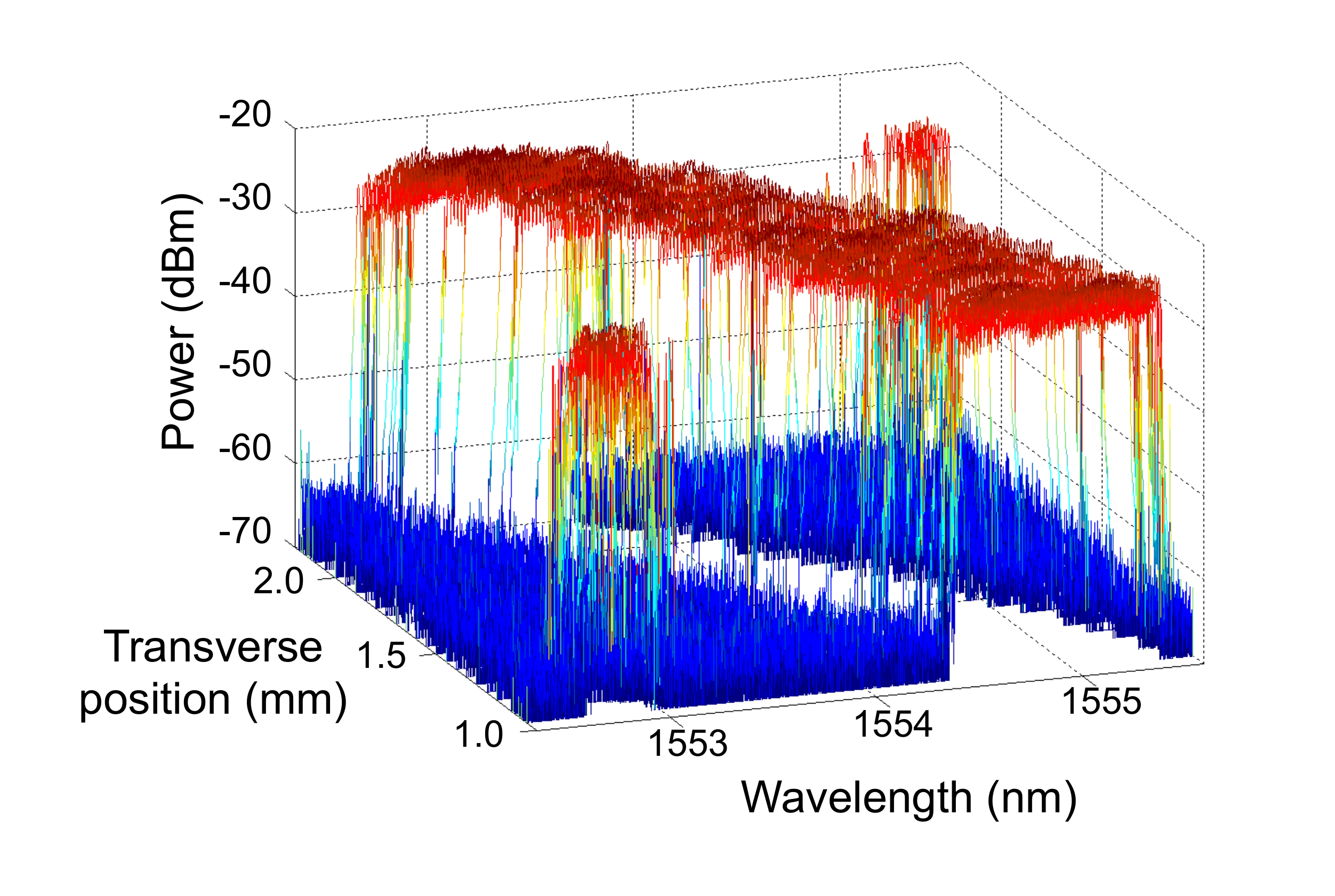}
\caption{Signal spectra recorded for different transverse positions of the crystal. The input pump power is 1 W and the poling period is $\Lambda = 30.25 \mu \textrm{m}$. The change of shape of the oscillation bandwidth is due to the change of the relative phase between the three back-reflected waves.}
\label{Fig05}
\end{figure}

The control of the parametric gain bandwidth allowed by the fine tuning of the phase difference between the three back-reflected waves is illustrated by the experimental results of Fig.\,\ref{Fig05}. The different spectra reproduced in this figure were obtained by adjusting the transverse position of the nonlinear crystal while remaining inside the same poling period region ($\Lambda = 30.25\, \mu \textrm{m}$). Like in Fig.\,\ref{Fig04}, each spectrum is obtained by scanning both mirror longitudinal positions. One can clearly see that the association of a wedge angle and an achromatic mirror M$_3$ allows some parametric gain shaping in the cw regime \cite{Bjorkholm1970,Hardy2010}.

\section{Management of thermal effects and tunability}

\begin{figure}[h!]
\centering
\includegraphics[width=0.6\columnwidth]{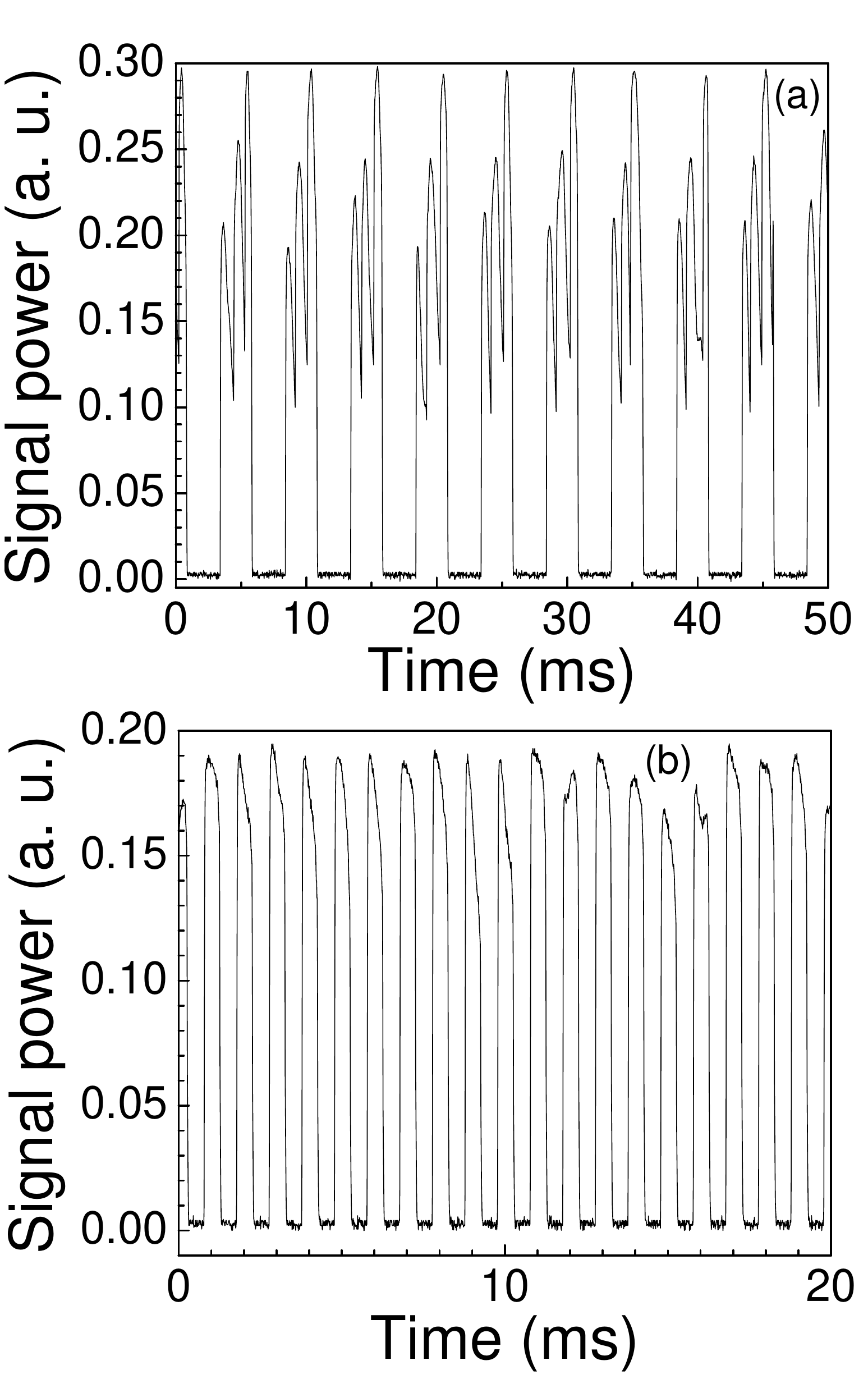}
\caption{Time evolution of the OPO signal power when the pump is chopped at 200~Hz (a) and 1~kHz (b). The input pump power is 1.2~W and the poling period is $\Lambda = 30.75\,\mu \textrm{m}$. In this configuration the radius of curvature of mirror M$_3$ is 30 mm and the idler peak power is equal to 19~mW.}
\label{Fig06}
\end{figure}

As shown by the results of Fig.\,\ref{Fig04}, thermal effects play a more important role in cw regime than in pulsed regime \cite{Hardy2011,Barria2013}. Indeed, in cw regime, we observe that tuning the position of mirror M$_1$ or mirror M$_3$ does not allow to tune the frequencies of the OPO in a controllable and reproducible manner. In order to reduce the influence of these thermal effects while keeping the same peak power for the OPO, we chose to mechanically chop the pump beam at the entrance of the OPO (for a similar technique applied to SROs see Refs. \cite{Melkonian2007,Samanta2007,Samanta2008}). In these conditions, we record the evolution of the OPO signal power versus time.  Fig.\,\ref{Fig06} reproduces such evolutions for two different chopping frequencies. Fig.\,\ref{Fig06}(a), obtained for a chopper frequency of 200~Hz, clearly shows that the OPO exhibits several (typically three) mode hops during each oscillation window lasting 2.5~ms. This indicates that the rise time for thermal effects in the nonlinear crystal is of the order of 1~ms. Consequently, if we chop the pump beam at a frequency faster than 1~kHz, we do no longer allow thermal effects to induce mode hops. This is evidenced by the plot of Fig.\,\ref{Fig06}(b), which clearly shows that the OPO does no longer exhibit any mode hop. 

\begin{figure}[h!]
\centering
\includegraphics[width=0.6\columnwidth]{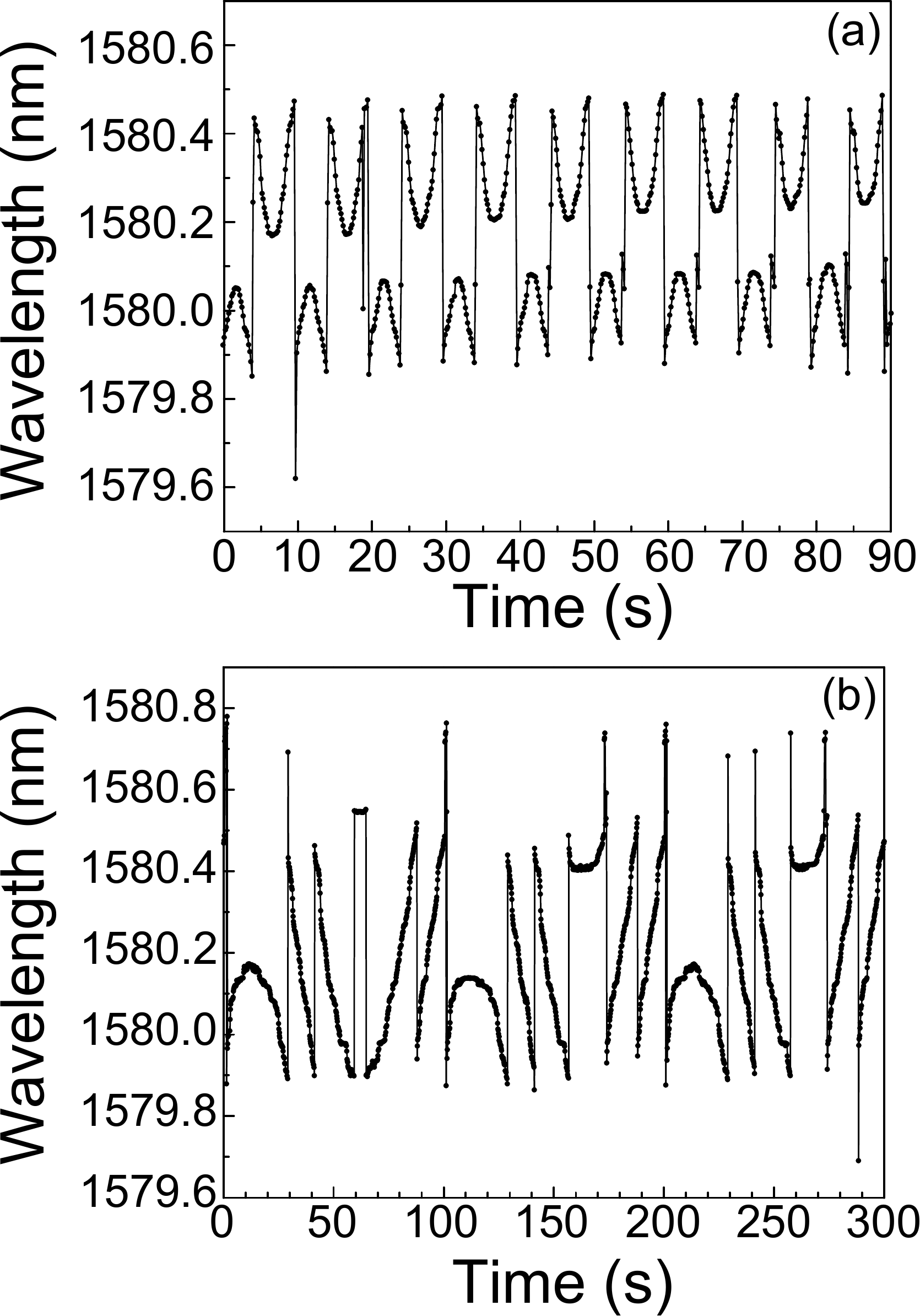}
\caption{Time evolution of the signal wavelength when the longitudinal position of mirror M$_1$ (a) or M$_3$ (b) is scanned at 100~mHz and 10~mHz, respectively, while the pump power is chopped at 4~kHz. The input pump power is 1.5~W and the poling period is $\Lambda = 30.50\,\mu \textrm{m}$. In this configuration the radius of curvature of mirror M$_3$ is 30 mm and the idler peak power is equal to 34~mW.}
\label{Fig07}
\end{figure}

Such chopping frequencies larger than 1~kHz are optimal for photoacoustic detection of trace gases \cite{Barria2013}. We have thus characterized the tuning capabilities of our nested cavity OPO while the pump is chopped at 4~kHz. The results of Fig.\,\ref{Fig07} illustrate this tunability. Fig.\,\ref{Fig07}(a) corresponds to the evolution of the OPO wavelength when a sinusoidal voltage at 100~mHz is applied to the PZT carrying mirror M$_1$. In this situation, only the idler cavity length is modulated. This figure shows that in the presence of the chopper, the OPO behaves like the preceding examples of nested cavity OPOs that were operated in the ns regime \cite{Hardy2011}. 

Indeed, tuning the length of the idler cavity allows to explore the parametric gain bandwidth by successive mode hops. When the voltage is applied to the PZT carrying mirror M$_3$ [see Fig.\,\ref{Fig07}(b)], both idler and signal cavity lengths are modulated. This explains the more complicated shape of the wavelength evolution. However, here also the wavelength evolution is reproducible. This shows that the present architecture is compatible with tuning strategies earlier developed in the nanosecond and microsecond regimes \cite{Barria2013,Hardy2012}, while being optimized for photoacoustic detection modulation frequencies.

\section{Conclusion}

In conclusion, we demonstrated a continuous-wave doubly resonant optical parametric oscillator based on a nested linear cavity. Single longitudinal mode operation was obtained with low threshold power ($<$ 500mW) compatible with diode laser pumping. The control of the oscillation bandwidth resulting from the fine adjustment of the relative phase between the back-reflected waves, obtained by the beveled crystal end-face, was obtained in continuous-wave regime. Moreover, the thermal effects inherent to the cw operation regime have been shown to be quite easy to manage by chopping the pump laser at frequencies optimized for photoacoustic absorption spectroscopy. This opens the possiblity to use such a cw DRO for trace gas detection and spectroscopy with a signal-to-noise ratio considerably increased with respect to pulsed solution, while maintaining a compact and robust architecture and opening the way to pumping by compact, efficient, and single-frequency diode lasers.

\section*{References}

\bibliography{mybibfileReRevised}

\end{document}